\documentclass[prb,twocolumn]{revtex4}
\usepackage{graphicx,amsmath,amssymb,amsfonts,latexsym,color,dcolumn,bm,mathtools}
\usepackage{verbatim}
\usepackage{epstopdf}
\usepackage[colorlinks=true,hidelinks]{hyperref}

\providecommand{\moy}[1]{\langle #1 \rangle}

\definecolor{blue}{rgb}{0.223,0.223,0.667}
\definecolor{red}{rgb}{0.7,0,0}

\begin{document}

\title{Quantum nondemolition measurement of a nonclassical state of a massive object}

\author{F. Lecocq, J. B. Clark,  R. W. Simmonds, J. Aumentado, J. D. Teufel}

\affiliation{National Institute of Standards and Technology, 325 Broadway, Boulder, CO 80305, USA}

\date{\today}

\maketitle

\textbf{While quantum mechanics exquisitely describes the behavior of microscopic systems, one ongoing challenge is to explore its applicability to systems of larger size and mass.  Unfortunately, quantum states of increasingly macroscopic objects are more easily corrupted by unintentional measurements from the classical environment. Additionally, even the intentional measurements from the observer can further perturb the system\cite{Clerk2010}. In optomechanics\cite{Aspelmeyer2013}, coherent light fields serve as the intermediary between the fragile mechanical states and our inherently classical world by exerting radiation pressure forces and extracting mechanical information. Here we engineer a microwave cavity optomechanical system\cite{Teufel2011} to stabilize a nonclassical steady-state of motion while independently, continuously, and nondestructively monitoring it. By coupling the motion of an aluminum membrane to two microwave cavities, we separately prepare and measure a squeezed state of motion\cite{Kronwald2013}. We demonstrate a quantum nondemolition (QND) measurement\cite{Braginski1980,Clerk2008,Suh2014} of sub-vacuum mechanical quadrature fluctuations.  The techniques developed here have direct applications \cite{Metcalfe2014} in the areas of quantum-enhanced sensing\cite{Giovannetti2004} and quantum information processing, and could be further extended to more complex quantum states\cite{Leghtas2015}.} 

Reflecting light off a mechanical object induces a momentum transfer, allowing one to control and  measure the mechanical state. When the photon scattering rate exceeds the phonon decoherence rate the mechanical system becomes more strongly coupled to the photon reservoir than to its own thermal environment. This regime is usually obtained by embedding a mechanical resonator into an electromagnetic cavity to increase the interaction strength per photon\cite{Aspelmeyer2013}. Additionally, the cavity filters the density of states available for the scattered photons, allowing control over the ratio of Stokes and anti-Stokes scattering rates. Importantly, the nature of the optomechanical interaction implies that the light field interacts with both mechanical quadratures, with fundamental consequences on the mechanical state preparation and measurement. On one hand, the precision on the simultaneous measurement of both mechanical quadratures is limited by the Heisenberg's uncertainty principle\cite{Purdy2013b,Teufel2015}. On the other hand, the state preparation via sideband cooling exploits the coherent exchange of the cavity and mechanical state and is therefore limited by the statistics of the classical light field\cite{Teufel2011b,Chan2011,Palomaki2013}.

Both limitations can be overcome using polychromatic coherent light. One can address and manipulate each mechanical quadrature differently by engineering interference processes between their couplings to the cavity quadratures. More specifically, two-drive schemes can be used to design a single quadrature measurement of the mechanical oscillator, known as \textit{backaction evading}\cite{Braginski1980,Clerk2008,Suh2014}. The scheme fulfills the requirement for a QND measurement, which is an important tool for the tomographic reconstruction of arbitrary quantum states. A similar scheme was proposed by Kronwald \textit{at al}\cite{Kronwald2013} to prepare a mechanical squeezed state, following an analogous idea formulated for trapped ions\cite{Cirac1993,Kienzler2015}. In this work, we simultaneously implement these two schemes in a single microwave optomechanical system to perform the tomographic measurement of a squeezed state of a macroscopic mechanical oscillator.

\begin{figure*}
	\includegraphics[scale=1]{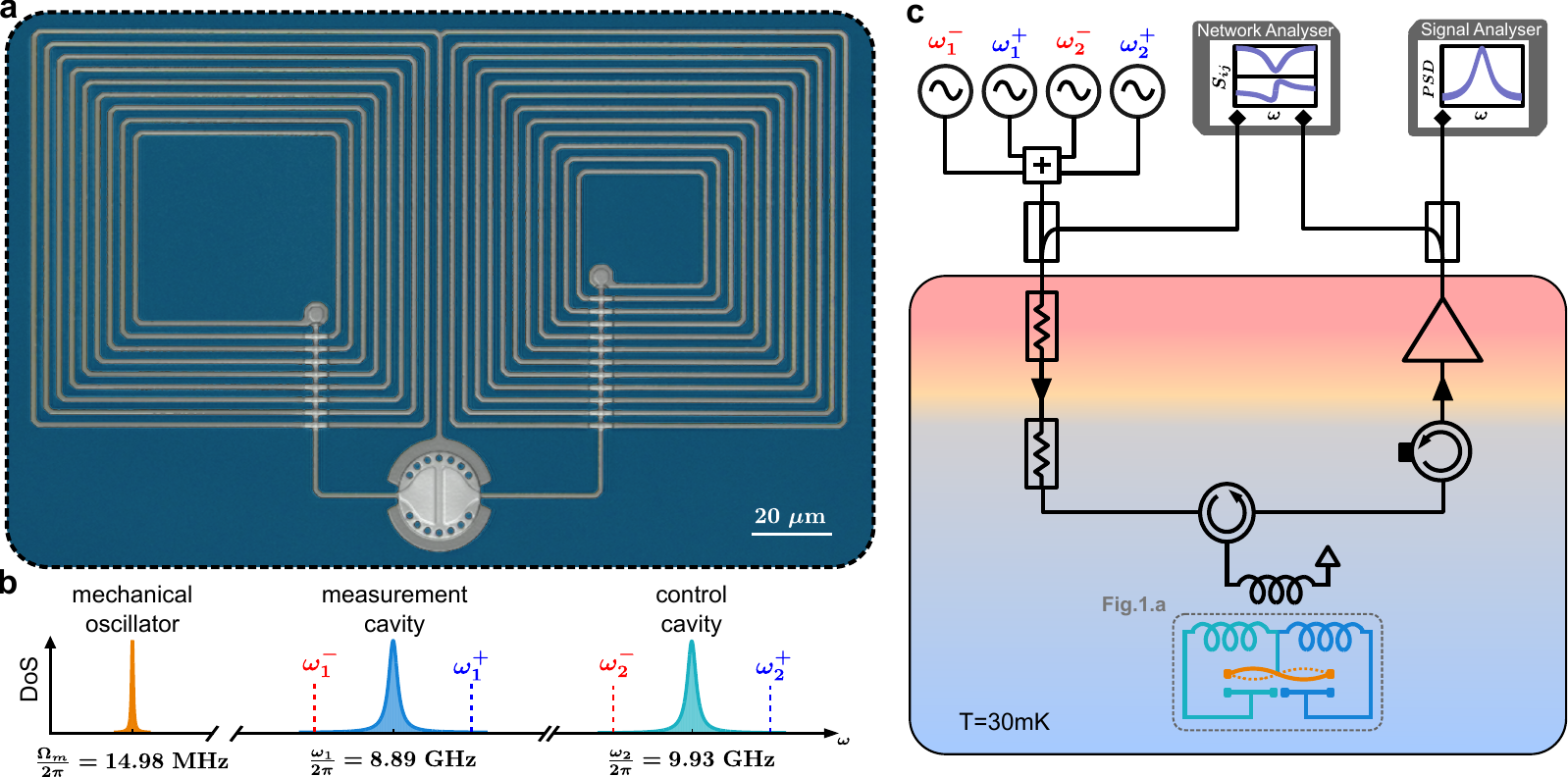}
	\caption{\textbf{Device description and experimental setup}. \textbf{a}, False-colour optical micrograph of the aluminum device (in grey) on a sapphire substrate (blue). Centered at the bottom of the micrograph is a mechanically compliant vacuum gap capacitor. The capacitor's electrode is split into two plates, each shunted by a different coil inductor, giving rise to two microwave LC resonances. \textbf{b}, Frequency space diagram. Except for the mechanical linewidth, all the frequencies and linewidths are to scale. The microwave cavities have Lorentzian densities of states of width $\kappa_1/2\pi=1.7~\mathrm{MHz}$ and $\kappa_2/2\pi=2.1~\mathrm{MHz}$, centered at $\omega_1/2\pi=8.89~\mathrm{GHz}$ and $\omega_2/2\pi=9.93~\mathrm{GHz}$. These resonance frequencies are tuned by the motion of the top plate of the capacitor, at the mechanical frequency $\Omega_m/2\pi=14.98~\mathrm{MHz}$. The red and blue dashed lines indicate the four mechanical sideband frequencies at which the cavities are driven, $\omega_{1,2}^{\pm}=\omega_{1,2}\pm\Omega_m$.  \textbf{c}, The circuit is placed on the cold stage of a cryogenic refrigerator (base temperature $T=30~\mathrm{mK}$). Up to four strong microwave drives and one weak microwave probe are inductively coupled to both cavities via a single port. The reflected signals and the noise emitted by the cavities are cryogenically amplified and demodulated at room temperature.
		\label{fig1}}
\end{figure*}

Consider a cavity optomechanical system where the position of a mechanical resonator of frequency $\Omega_m$ tunes the resonance frequency $\omega_c$ of an electromagnetic cavity\cite{Aspelmeyer2013}. Two drives are applied to the cavity, at both mechanical sidebands $\omega^{\pm}=\omega_{c}\pm\Omega_m$. The strength of each drive can be parametrized by its scattering rate $\Gamma^{\pm}=4g_0^2n^{\pm}/\kappa$, where $g_0$ is the vacuum optomechanical coupling rate, $\kappa$ is the cavity  linewidth and $n^{\pm}$ is the number of intra-cavity photons induced by each drive. Assuming a mechanical relaxation rate $\Gamma_m$ and the condition $\Gamma_m,\Gamma^{\pm}\ll\kappa\ll\Omega_m$, one can write the relations between the mechanical quadrature amplitudes, $\hat{X}_1$ and   $\hat{X}_2$, and the amplitude and phase quadrature of the cavity fields, $\hat{A}$ and $\hat{\varphi}$, reading\cite{Kronwald2013} :

\begin{flalign}
& 	\moy{\hat{X}_1^{2}}= \frac{\Gamma_m\moy{\hat{X}_{th}^{2}}+\left(\sqrt{\Gamma^-}-\sqrt{\Gamma^+}\right)^2\moy{\hat{\varphi}^{2}}}{\Gamma_m+\Gamma^--\Gamma^+} \label{VarX} \\
& 	\moy{\hat{X}_2^{2}}= \frac{\Gamma_m\moy{\hat{X}_{th}^{2}}+\left(\sqrt{\Gamma^-}+\sqrt{\Gamma^+}\right)^2\moy{\hat{A}^{2}}}{\Gamma_m+\Gamma^--\Gamma^+} \label{VarY}
\end{flalign}

Here $\moy{X_{th}^{2}}=2n_m^{th}+1$ is the variance of the mechanical quadratures for an equilibrium thermal occupancy $n_m^{th}$, and $\moy{\hat{\varphi}^{2}}=\moy{\hat{A}^{2}}=1$ are the variances of the cavity quadratures for an ideal coherent state. For $\Gamma^+=0$, corresponding to driving only the lower sideband, one recovers the sideband cooling limit and at high scattering rate, $\Gamma^-\gg\Gamma_m$, each quadrature of the mechanics is cooled to the cavity quadratures, $\moy{\hat{X}_1^{2}}=\moy{\hat{\varphi}^{2}}$ and $\moy{\hat{X}_2^{2}}=\moy{\hat{A}^{2}}$. Another limit is $\Gamma^+=\Gamma^-$, corresponding to driving symmetrically the upper and lower sidebands. This is the case of a drive on resonance with the cavity whose amplitude is modulated at a mechanical frequency, performing a QND measurement of the mechanical quadrature $\hat{X}_1$. Indeed, under these conditions, Eqs.\ref{VarX} and \ref{VarY} read $\moy{\hat{X}_1^{2}}=\moy{\hat{X}_{th}^{2}}$ and $\moy{\hat{X}_2^{2}}=\moy{\hat{X}_{th}^{2}}+(4\Gamma^-/\Gamma_m)\moy{\hat{A}^{2}}$. The $\hat{X}_1$ quadrature is unaffected by the measurement and the radiation pressure shot noise backaction is placed on the orthogonal quadrature $\hat{X}_2$. Finally the preparation of a squeezed state occurs in the intermediate regime, $\Gamma^+<\Gamma^-$. The mechanical mode is coupled, at a reduced rate $\Gamma^--\Gamma^+$, to an effectively squeezed microwave bath, whose minimum variance is $(\sqrt{\Gamma^-}-\sqrt{\Gamma^+})^2/(\Gamma^--\Gamma^+)<1$.

In order to separately prepare and read out a mechanical state, we engineer a microwave optomechanical system where a single mechanical mode is coupled to two microwave cavities. The experimental setup is shown in Fig.\ref{fig1}. The circuit, made out of aluminum on a sapphire substrate, consists of a central vacuum gap capacitor shunted by two coil inductors \cite{Cicak2010,Teufel2011}. The bottom plate of the capacitor is split to create two cavity resonances, $\omega_1/2\pi=8.89~\mathrm{GHz}$ and $\omega_2/2\pi=9.93~\mathrm{GHz}$, named respectively the ``measurement cavity'' and ``control cavity''. The top plate of the capacitor is mechanically compliant, with a first harmonic mode of motion resonating at $\Omega_m/2\pi=14.98~\mathrm{MHz}$. Its motion tunes the resonance of both microwave cavities, with respective vacuum optomechanical couplings $g_1/2\pi=145~\mathrm{Hz}$ and $g_2/2\pi=170~\mathrm{Hz}$. Operated at a temperature of $T=30~\mathrm{mK}$, the equilibrium mechanical thermal occupancy is $n_m^{th} = 42$ phonons and the mechanical relaxation rate is $\Gamma_m/2\pi=9.2~\mathrm{Hz}$. Both microwave cavities are strongly overcoupled to a single measurement port, setting their linewidths to $\kappa_1/2\pi=1.7~\mathrm{MHz}$ and $\kappa_2/2\pi=2.1~\mathrm{MHz}$. This coupling ensures that internal dissipations contribute by less than $5\%$ to the total linewidths, while maintaining a strongly resolved sideband regime, $\Omega_m/\kappa_{1,2}>7$. It also thermalizes the cavities to the shot-noise-limited input fields, maintaining throughout this work a thermal cavity occupancy well under our measurement noise floor, $n_c^{th}<0.1$.

\begin{figure}
	\includegraphics[scale=1]{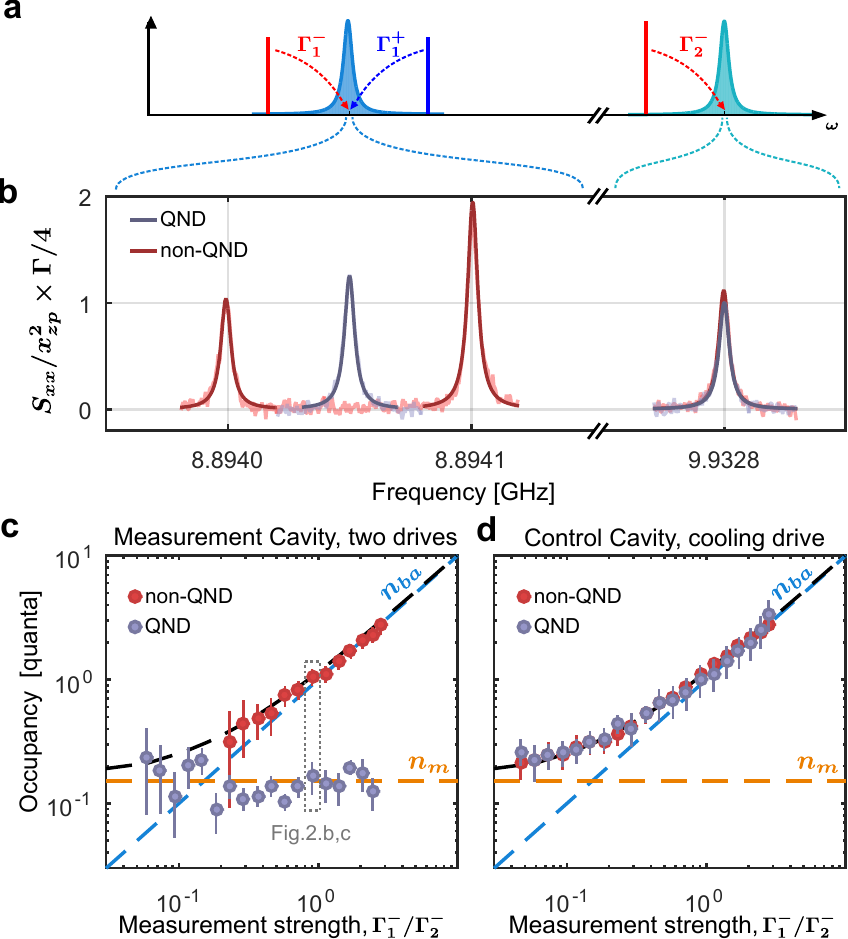}
	\caption{\textbf{Quantum nondemolition measurement}. \textbf{a}, Measurement schematic. A cooling drive of strength $\Gamma_{2}^{-}=2\pi\times4.87~\mathrm{kHz}=529\times\Gamma_m$ is applied on the lower mechanical sideband of the control cavity ($\omega_{2}^{-}=\omega_{2}-\Omega_m$). Two drives of equal strength $\Gamma_{1}^{-}=\Gamma_{1}^{+}$ are applied close to the mechanical sidebands of the measurement cavity. Their frequencies can be tuned to $\omega_{1}^{\pm}=\omega_{1}\pm\Omega_m$ to perform a single mechanical quadrature measurement (QND, in grey in \textbf{b-d}), or detuned by many mechanical linewidths away from that optimum to measure both mechanical quadratures (non-QND, in red in \textbf{b-d}). \textbf{b}, Mechanical noise spectrum (normalized, background subtracted), for $\Gamma_{1}^{-}/\Gamma_{2}^{-}=0.9$. \textbf{c-d}, Mechanical occupancy extracted from the measured spectra of the two-drives measurement (\textbf{c}) and cooling drive (\textbf{d}), for both the non-QND case (in red) and the QND case (in grey), as a function of the measurement strength $\Gamma_{1}^{-}/\Gamma_{2}^{-}$.
		\label{fig2}}
\end{figure}

We start by describing the QND measurement of the mechanical oscillator, cooled close to its ground state, in Fig.\ref{fig2}. A cooling drive of strength $\Gamma_{2}^{-}=2\pi\times4.87~\mathrm{kHz}=529\times\Gamma_m$ is applied at the lower mechanical sideband of the control cavity, $\omega_{2}^{-}=\omega_{2}-\Omega_m$, leading to a reduced mechanical thermal occupancy $n_m$. Simultaneously, two drives of equal strength, $\Gamma_{1}^{-}=\Gamma_{1}^{+}$, are applied close to the mechanical sidebands of the measurement cavity, acting back on the mechanical oscillator and increasing the total occupancy to $n_m^{tot}=n_m+n_{ba}$ where $n_{ba}=\Gamma_{1}^{-}/\Gamma_{2}^{-}$. Their frequencies can be optimally tuned to $\omega_{1}^{\pm}=\omega_{1}\pm\Omega_m$ to perform a single mechanical quadrature measurement (QND, in grey in Fig.\ref{fig2}b-d), or detuned by many mechanical linewidths away from that optimum to measure both mechanical quadratures (non-QND, in red in Fig.\ref{fig2}b-d). By monitoring the driven responses of both cavities\cite{Weis2010,Teufel2011,Safavi-Naeini2011} we tune very precisely the strength of each drive and measure all the mode frequencies and decay rates [Supp. Inf.]. We then acquire the noise power emitted by both cavities. In Fig.\ref{fig2}b, we fix the  measurement rate to $\Gamma_{1}^{-}=\Gamma_{1}^{+}=0.9\times\Gamma_{2}^{-}$, and show the measured spectra, normalized to mechanical units [Supp. Inf.]. 

In the non-QND case, each drive measures both mechanical quadratures, and the noise power of the thermomechanical sidebands are proportional to $n_m^{tot}$ and $n_m^{tot}+1$ for the anti-Stokes and Stokes scattering, respectively\cite{Safavi-Naeini2012}. Note that this sideband asymmetry\cite{Weinstein2014,Lecocq2015} provides a primary calibration of the y-axis, in good agreement with the independently measured coupling strengths $g_{1}$ and $g_{2}$. From each sideband, we extract the same mechanical occupancy $n_m^{tot}$, shown as a function of measurement strength in Fig.\ref{fig2}c-d. The measured quantum backaction scales ideally with the measurement strength, and we can extrapolate a mechanical thermal occupancy of $n_m=0.15\pm0.05$. 

We now tune the frequency of the measurement drives to the QND case. As shown in Fig.\ref{fig2}b, the noise sideband of the cooling drive is unchanged. Indeed, that drive still measures both mechanical quadratures, accessing the same total mechanical occupancy as in the non-QND measurement (see Fig.\ref{fig2}d). To the contrary, on the measurement cavity, the mechanical sidebands of each drive interfere with each other when brought into the QND case, leaving a single Lorentzian noise peak proportional to the variance of a single mechanical quadrature given by $\moy{\hat{X}_1^{2}}=2n_m+1$. The backaction has been evaded and placed on the orthogonal quadrature, conserving the total mechanical occupancy (see Eq.\ref{VarX} and Eq.\ref{VarY}). As expected the measured mechanical thermal occupancy is constant as a function of the measurement strength, and quantitatively agrees with the occupancy inferred in the non-QND case. At the measurement strength of $\Gamma_{1}^{-}/\Gamma_{2}^{-}=2.44$ we measure an evasion of the quantum measurement backaction by more than $13~\mathrm{dB}$. This demonstrates a QND measurement of a single mechanical quadrature at a rate much faster than the mechanical decoherence rate.

\begin{figure*}
	\includegraphics[scale=1]{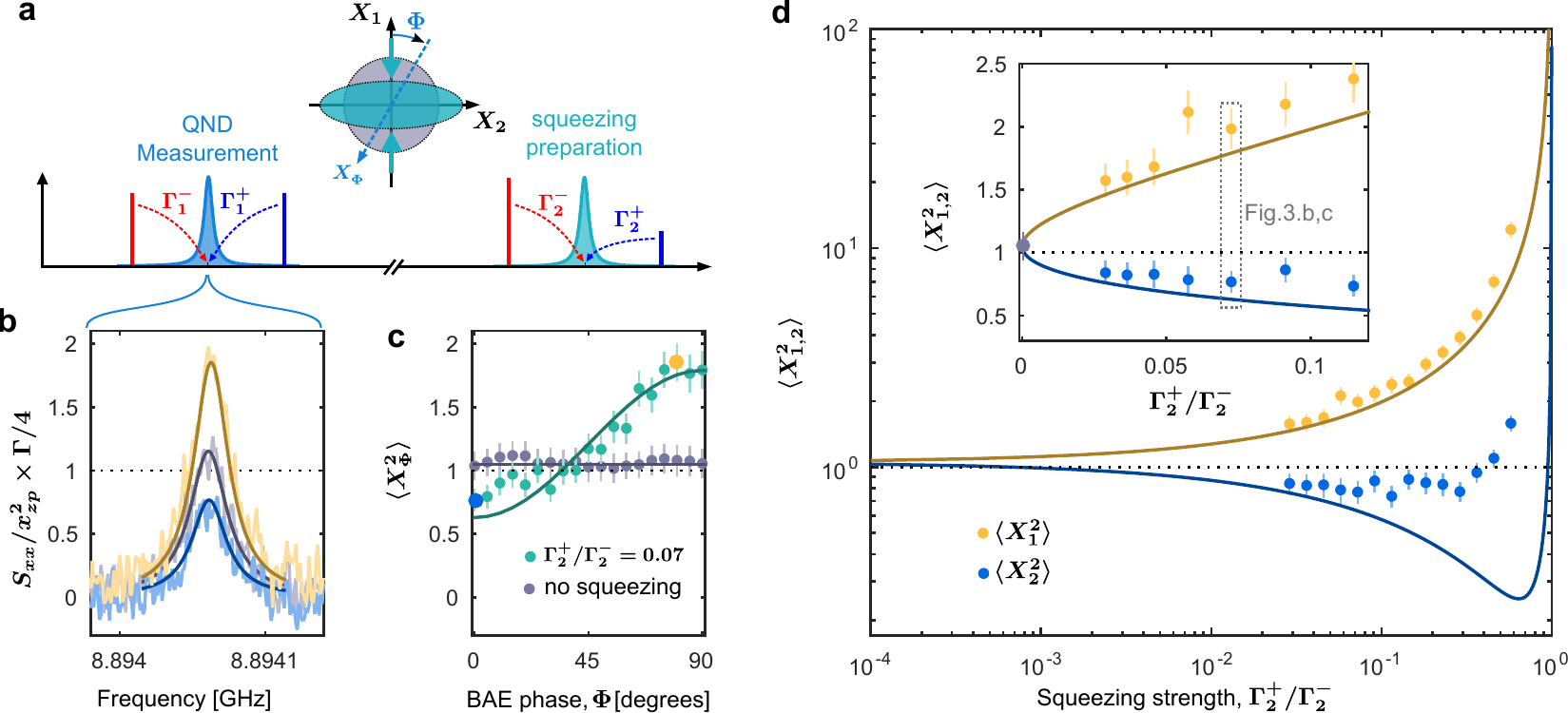}
	\caption{\textbf{Tomography of a mechanical squeezed state}. \textbf{a}, Measurement schematic. A pair of drives at $\omega_{2}^{\pm}=\omega_{2}\pm\Omega_m$ cool the mechanical mode to a squeezed bath, while a pair of drives at $\omega_{1}^{\pm}=\omega_{1}\pm\Omega_m$ measure the generalized mechanical quadrature $\hat{X}_\Phi$, given by their relative and tunable phase $\Phi$, allowing for the tomographic measurement of the squeezed state. \textbf{b}, Normalized noise spectra of the QND measurement (background subtracted), for the squeezed and anti-squeezed quadratures ($\Gamma_{2}^{-}/\Gamma_{2}^{+}=0.07$) in blue and yellow respectively, compared to a spectrum measured without squeezing ($\Gamma_{2}^{+}=0$) in grey. \textbf{c}, Measured quadrature variances as a function of phase for  $\Gamma_{2}^{+}=0$ in grey and $\Gamma_{2}^{-}/\Gamma_{2}^{+}=0.07$ in green. The spectra in \textbf{b} correspond to the blue and yellow dots. \textbf{d}, Squeezed and anti-squeezed quadrature variances, respectively in blue and yellow, as a function of the squeezing strength $\Gamma_{2}^{-}/\Gamma_{2}^{+}=0.07$. The solid lines are the theoretical prediction with no free parameters.  The inset is a zoom-in for the low squeezing strength data, on a linear scale, where we can place the variance measured without squeezing (grey dot). In all figures, the black dotted line is the vacuum limit.
		\label{fig3}}
\end{figure*}

This measurement scheme allows us to perform the tomography of the mechanical state, described in Fig.\ref{fig3}. Indeed, we can control the generalized mechanical quadrature being measured, $\hat{X_{\Phi}}$, by simply rotating the relative phase between the measurement drives. As we expect the state to be Gaussian, the measurement of the second moment of the noise is sufficient to reconstruct its tomogram. A mechanical state prepared by simple sideband cooling is expected to have equal variances for each quadrature. In Fig.\ref{fig3}c, we show data for an increased cooling strength  $\Gamma_{2}^{-}=2\pi\times15.11~\mathrm{kHz}=1643\times\Gamma_m$ and a measurement strength $\Gamma_{1}^{-}/\Gamma_{2}^{-}=0.48$. As expected the results of the QND measurements are phase independent. We measure a mechanical occupancy $n_m < 0.1 $, demonstrating the QND measurement of a highly pure Gaussian state.

We now apply this same tomographic measurement to verify the preparation of a non-classical state of motion. We prepare a squeezed state by adding a drive of strength $\Gamma_{2}^{+}$ at the upper mechanical sideband of the control cavity ($\omega_{2}^{+}=\omega_{2}+\Omega_m$). Again, as a squeezed state is still Gaussian, we simply measure the variance of the mechanical quadrature $\hat{X_{\Phi}}$ as a function of the measurement phase to fully characterize the state, as shown in Fig.\ref{fig3}c for a squeezing strength $\Gamma_{2}^{+}/\Gamma_{2}^{-}=0.07$. We resolve a minimum quadrature variance below vacuum, $\moy{\hat{X}_1^2}=0.78\pm0.08$. The spectra corresponding to the measurement of the squeezed and anti-squeezed quadratures are shown in Fig.\ref{fig3}b. Finally, in Fig.\ref{fig3}.d, we plot the variance of the squeezed and anti-squeezed quadratures, $\moy{\hat{X}_1^2}$ and $\moy{\hat{X}_2^2}$ respectively, as a function of the squeezing strength. The solid lines in Fig.\ref{fig3}c,d are theoretical predictions from Eq.\ref{VarX} and \ref{VarY} without free parameters, in reasonable agreement with the data at low squeezing strength. To further understand the deviation between data and theory at high squeezing strength, future experiments will investigate the effects of frequency noise of the cavities and the mechanical oscillator, residual thermal cavity occupancy or phase instability of the drives. We emphasize that acquiring all the moments of the noise emitted by the measurement would allow us to reconstruct an arbitrary quantum state without making a Gaussian assumption\cite{Mallet2011,Eichler2011}.

Looking forward, the introduction of stronger non-linearities, combined with reservoir engineering, would enable the preparation of more complex quantum states \cite{Leghtas2015}, further motivating the use of mechanical systems as ultra-sensitive detectors and quantum memories\cite{Metcalfe2014}. Additionally, the inherent non-linearity of optomechanical cavities can act as a nearly-ideal mixing element, opening routes for innovative types of amplification\cite{Metelmann2014}, frequency conversion\cite{Safavi-Naeini2011a}, and non-reciprocal behavior\cite{Metelmann2015}.

Note: While preparing the manuscript we became aware of other work using a similar method for the mechanical state preparation\cite{Wollman2015,Pirkkalainen2015}.


\end{document}


\title{Supplementary Information for ``Quantum nondemolition measurement of a nonclassical state of a massive object''}

\maketitle 

\textbf{Mechanical spectra and error estimation}

The mechanical state is extracted from the total microwave power spectral density, measured at room temperature with a spectrum analyser, following:

\begin{equation}
\frac{PSD^{\pm}\left[\omega\right]}{P_{d}}=\frac{16\eta^2\kappa^2 g_0^2}{\left[\kappa^2(1-2\eta)^2+4\Delta^2\right]\left[\kappa^2+4(\Delta\pm\Omega_m)^2)\right]}\left(\frac{4n\Gamma}{\Gamma^2+4\delta^2}+\frac{4n_{imp}}{\Gamma}\right)=A\frac{S_{xx}}{x_{zp}^2}+B
\end{equation}

where $PSD^{\pm}$ is the measured power spectral density at the upper or lower sideband of the drive in units of [W/Hz], $P_{d}$ is the measured pump power in units of [W], $\Delta = \omega_d - \omega_c$ is the drive detuning, $\delta = \omega - \omega_d\pm\Omega_m$ is the detuning around the mechanical sideband, $\omega_c$ is the microwave cavity frequency, $\Omega_m$ is the mechanical oscillator frequency, $\omega_d$ is the drive frequency, $\Gamma$ is the total mechanical linewidth, $\kappa = \kappa_{int} + \kappa_{ext}$ is the total cavity linewidth, $\eta=\kappa_{ext}/\kappa$ is the coupling factor, and $g_0$ is the optomechanical coupling strength. Here, $n_{imp}$ is the measurement noise floor expressed in unit of mechanical quanta, and $n = n_m^{tot}$ for the anti-Stokes scattering or $n = n_m^{tot}+1$ for the Stokes scattering.

From this measurement one can extract the mechanical displacement spectral density, $S_{xx}$, normalized by the variance of zero point fluctuation $x_{zp}^2 = \hbar/(2m\Omega_m)$, where $m$ is the mechanical oscillator's mass. In Fig.2.b and Fig.3.b, we show the quantity  $S_{xx}/x_{zp}^2\times\Gamma/4$, allowing one to read the value of $n$ off of the maximum value of the Lorentzian peak. We remove the noise floor $B$ for readability, as the noise floor differs slightly between the control and measurement cavity. The transduction factor $A$ and the linewidth $\Gamma$ are measured independently, as explained in the next sections.

In Figs.2 and 3, the error bar on the data correspond to the $90\%$ confidence intervals, obtained from standard error propagation. The main source of uncertainties in our experiment are due,  in order of greatest importance, to the error in the  Lorentzian fits of the spectra, to the uncertainty in the estimation of $g_{1,2}$ and of the cavities coupling factor $\eta_{1,2}$.
\\

\textbf{Calibrations and experimental details}

All the drives are filtered at room temperature and attenuated in the cryostat, ensuring that they are devoid of excess noise at the cavities frequencies\cite{Lecocq2015}.

\begin{figure*}
	\includegraphics[scale=1]{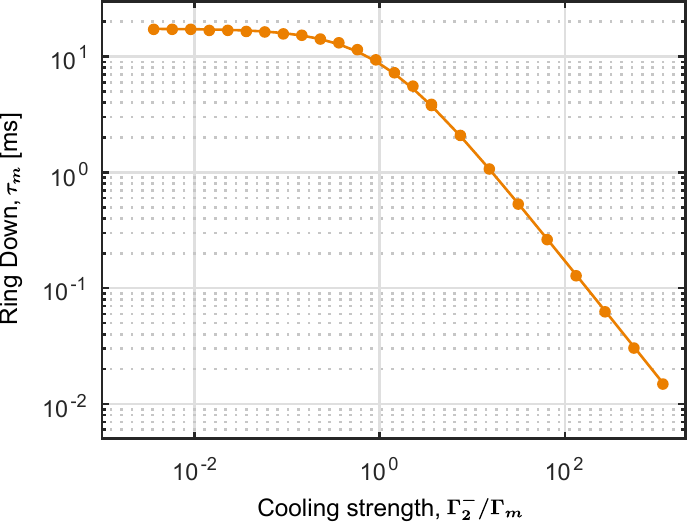}
	\caption{\textbf{Mechanical ring-down time}. We measure the mechanical ring down time $\tau_m$ as a function of the strength of a cooling drive on the control cavity, at a frequency $\omega_2^-=\omega_2-\Omega_m$, scattering light and damping the mechanical oscillator at the rate $\Gamma_2^-$. The solid line is the theoretical prediction from the total damping, $\Gamma_{tot}=\Gamma_m+\Gamma_2^-$. We measure an intrinsic mechanical ring-down time of $\tau_m=17.3~\mathrm{ms}$, corresponding to a relaxation rate $\Gamma_m=1/\tau_m=2\pi\times9.2~\mathrm{Hz}$.
		\label{figS1}}
\end{figure*}

The intrinsic mechanical relaxation rate $\Gamma_m$ is measured from the mechanical ring-down time $\tau_m$. In Fig.\ref{figS1}, we show the measured ring-down time as a function of the strength of a cooling drive on the control cavity, at a frequency $\omega_2^-=\omega_2-\Omega_m$, scattering light and damping the mechanical oscillator at the rate $\Gamma_2^-$. We measure an intrinsic ring-down time $\tau_m=17.3~\mathrm{ms}$, corresponding to a relaxation rate $\Gamma_m=1/\tau_m=2\pi\times9.2~\mathrm{Hz}$. This measurement also allows us to calibrate the cooling drive strength $\Gamma_2^-$.

The optomechanical coupling strengths for each cavity, $g_{1,2}$, are calibrated from a fridge temperature sweep like in Teufel, \textit{et al}\cite{Teufel2011b}. The measured values are confirmed by the measurement of the sideband asymmetry in Fig.2.

Finally the cavity frequencies and linewidths, the mechanical frequency, and every pump scattering rate, are independently measured from the driven responses (see next section).
\\

\textbf{Scattering Parameters}

Before measuring the spectra shown in Fig.2 and Fig.3, we measure the driven responses of the system in presence of all the pump drives. They contain all the information about our system, except for the bath temperatures, and represent an important calibration. With up to four drives, two cavity modes and a mechanical mode, and with correction coming from finite sideband-resolution, the analytic solutions for the driven responses can be cumbersome, if not intractable. However, they can easily be solved numerically from the linear coupled equations of motion, following methods like in Andrews, \textit{et al}\cite{Andrews2014} or Ranzani, \textit{et al}\cite{Ranzani2015}.

\begin{figure*}
	\includegraphics[scale=1]{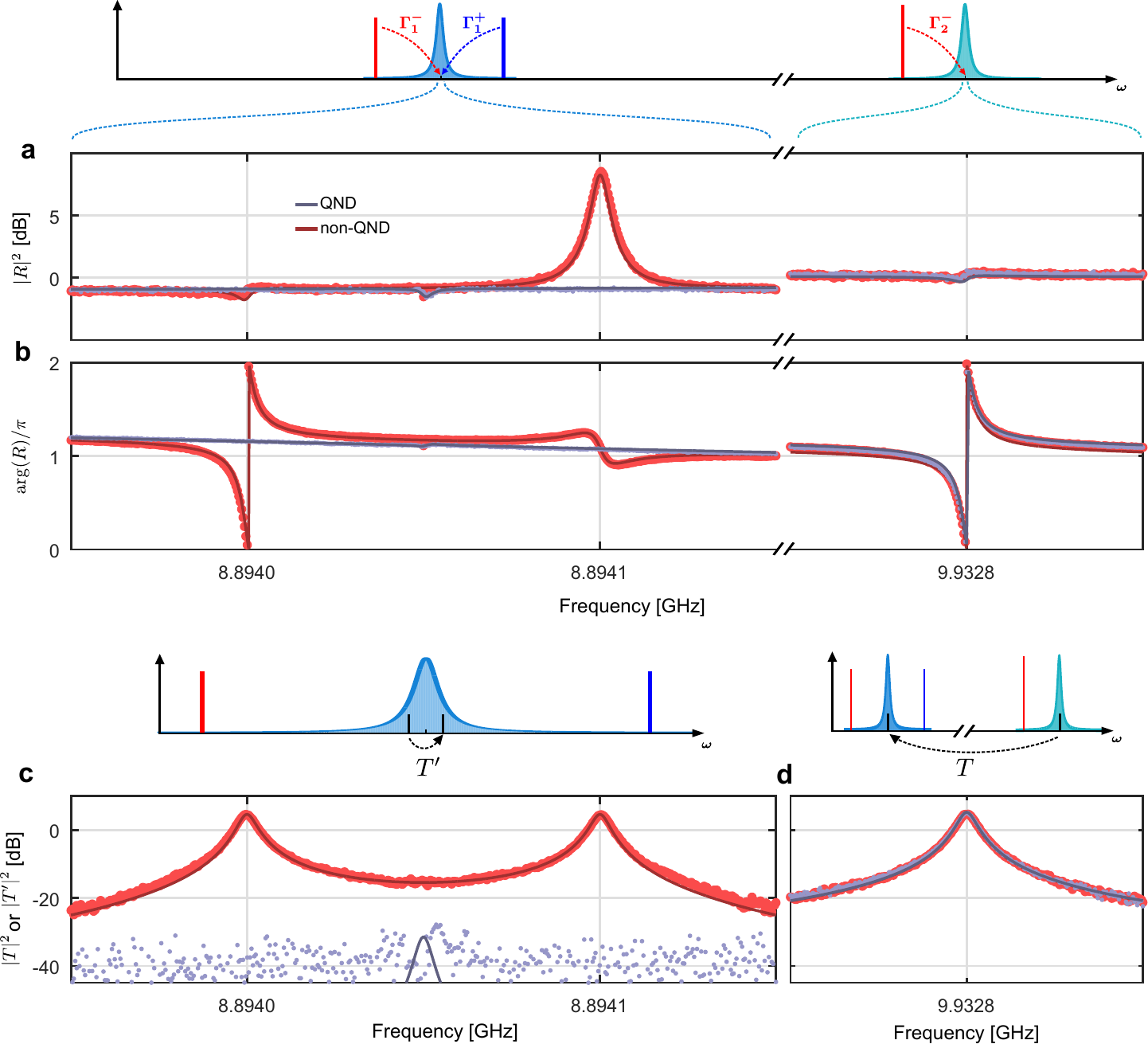}
	\caption{\textbf{Driven responses for the QND and non-QND measurement}. Data corresponding to the measurement setup in Fig.2: the pump scattering rates are $\Gamma_{2}^{-}=2\pi\times4.87~\mathrm{kHz}=529\times\Gamma_m$ and  $\Gamma_{1}^{-}=\Gamma_{1}^{+}=0.9\times\Gamma_{2}^{-}$. \textbf{a-b}, Reflection coefficient, $R$, around each cavity resonance. \textbf{c} Transmission from the upper mechanical sideband of the red measurement drive to the lower mechanical sideband of the blue measurement drive, as a function of the input frequency. A signal sent at $\omega$ is received at $\omega-\omega_{1}^{-}+\omega_{1}^{+}$. \textbf{d} Transmission from the upper mechanical sideband of the red cooling drive to the upper mechanical sideband of the red measurement drive, as a function of the input frequency. A signal sent at $\omega$ is received at $\omega-\omega_{2}^{-}+\omega_{1}^{-}$. All the solid lines are from the same numerical simulation as a function of pump frequency, allowing for a very precise measurement of the cavity frequencies and linewidths, the mechanical frequency, and every pump scattering rate.
		\label{figS2}}
\end{figure*}

\begin{figure*}
	\includegraphics[scale=1]{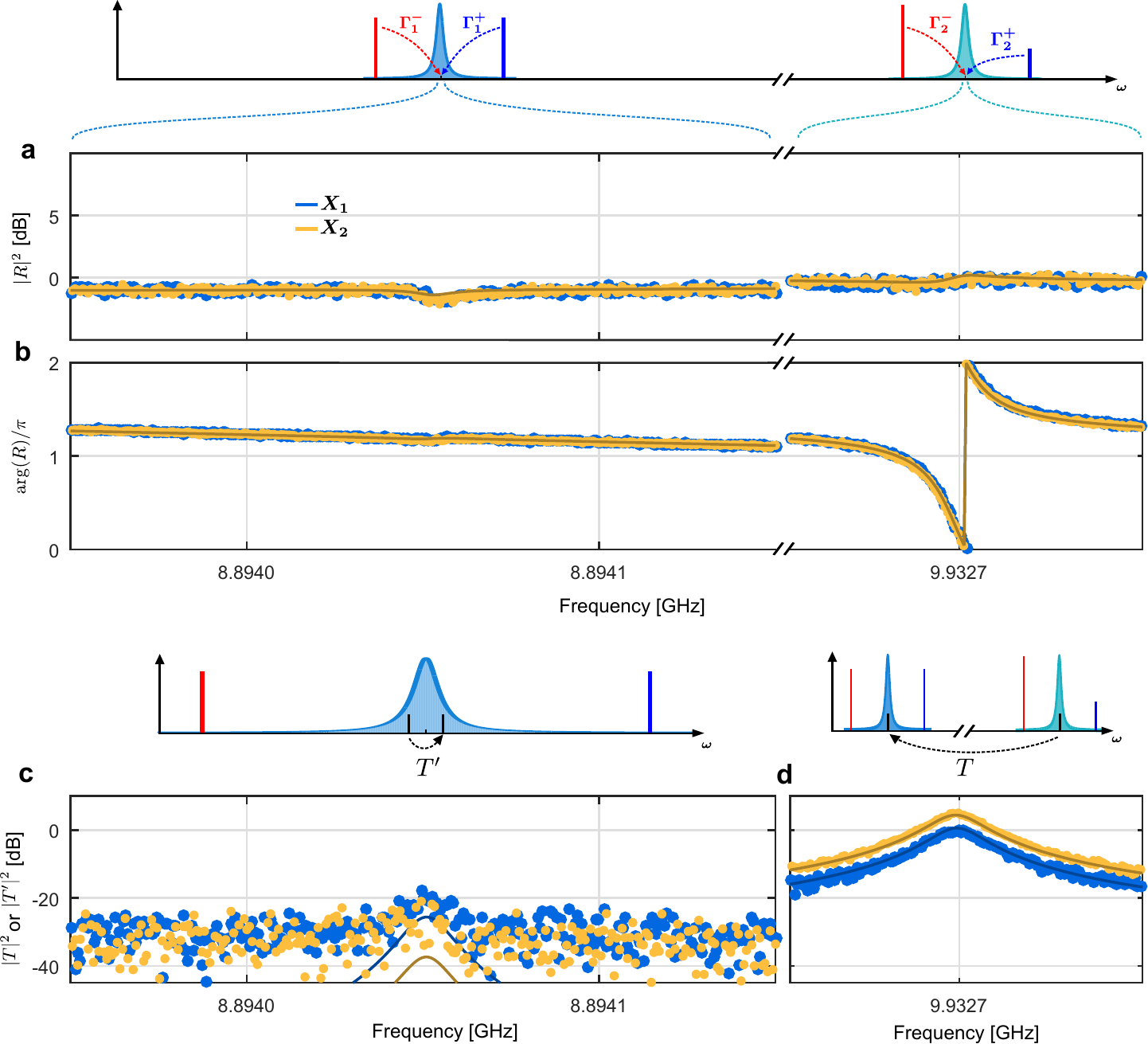}
	\caption{\textbf{Driven responses for the tomography of a mechanical squeezed state}. Data corresponding to the measurement setup in Fig.3: the pump scattering rates are $\Gamma_{2}^{-}=2\pi\times15.11~\mathrm{kHz}=1643\times\Gamma_m$, $\Gamma_{1}^{-}/\Gamma_{2}^{-}=0.48$ and $\Gamma_{2}^{+}/\Gamma_{2}^{-}=0.07$. \textbf{a-b}, reflection coefficient, $R$, around each cavity resonance. \textbf{c} Transmission from the upper mechanical sideband of the red measurement drive to the lower mechanical sideband of the blue measurement drive, as a function of the input frequency. A signal sent at $\omega$ is received at $\omega-\omega_{1}^{-}+\omega_{1}^{+}$. \textbf{d} Transmission from the upper mechanical sideband of the red cooling drive to the upper mechanical sideband of the red measurement drive, as a function of the input frequency. A signal sent at $\omega$ is received at $\omega-\omega_{2}^{-}+\omega_{1}^{-}$. All the solid lines are from the same numerical simulation as a function of pump phases, allowing for a very precise measurement of the cavity frequencies and linewidths, the mechanical frequency, and every pump scattering rate.
		\label{figS3}}
\end{figure*}

In Fig.\ref{figS2} and Fig.\ref{figS3} we show the driven responses corresponding respectively to the data shown in Fig.2 and Fig.3 in the main text. From the fit to these driven responses we extract the cavity frequencies and linewidths, the mechanical frequency, and every pump scattering rate.

